\documentclass[twocolumn,floatfix]{revtex4}
\usepackage{amssymb}
\usepackage{amsmath}
\usepackage{graphicx}

\usepackage{algorithm}
\usepackage{algorithmic}

\begin{document}

\title{Fast Community Identification by Hierarchical Growth}
\author{F. A. Rodrigues}
\author{G. Travieso}
\author{L. da F. Costa}
\affiliation{Institute of Physics of S\~{a}o Carlos. University of S\~{a}o Paulo\\
S\~{a}o Carlos, SP, Brazil, PO Box 369, 13560-970, phone +55 16 3373
9858,\\\{francisco,gonzalo,luciano\}@ifsc.usp.br}

\begin{abstract}
A new method for community identification is proposed which is founded
on the analysis of successive neighborhoods, reached through
hierarchical growth from a starting vertex, and on the definition of
communities as a subgraph whose number of inner connections is larger
than outer connections. In order to determine the precision and speed
of the method, it is compared with one of the most popular community
identification approaches, namely Girvan and Newman's
algorithm. Although the hierarchical growth method is not as precise
as Girvan and Newman's method, it is potentially faster than most
community finding algorithms.
\end{abstract}

 \maketitle

\section{Introduction}

Lying at the intersection between graph theory and statistical
mechanics, complex networks exhibit great generality, which has
allowed applications to many areas such as modeling of biological
systems~\cite{Barabasi:2004}, social
interactions~\cite{Newman:2003:Mixing,
Newman:2003,Gleiser:2003,Scott:2000} and information
networks~\cite{Faloutsos:1999,Albert:1999}, to cite just a
few~\cite{Barabasi:book}.

As this research area comes of age, a large toolkit is now available
to characterize and model complex networks (e.g.
surveys~\cite{Costa:survey,Boccaletti05,Barabasi:2002:survey,Newman:2003:survey,Barabasi:2002:survey}).
An important problem which has been subject of great interest recently
concerns the identification of modules of densely connected vertices
in networks, the so-called \emph{communities}.  These structures
result from interactions between the network components, defining
structural connecting patterns in social
networks~\cite{Arenas:2004,Gleiser:2003}, metabolic networks
\cite{Guimera:2005} as well as the worldwide air transportation
network \cite{Guimera:2005b}.

Despite the intense efforts dedicated to community finding, no
consensus has been reached on how to define
communities~\cite{Danon:2005}.  Radichi et
al.~\cite{Radicchi:2004} suggested the two following definitions.
In a strong sense, a subgraph is a community if all of its
vertices are more intensely connected one another than with the
rest of the network.  In a weak sense, a subgraph corresponds to a
community whenever the number of edges inside the subgraph is
larger than the number of connections established with the
remainder of the network.

Along the last few years, many methods have been proposed for
community identification based on a variety of distinct approaches
such as: (i) link removal, as used by Girvan and
Newman~\cite{Girvan:2003} and Radicchi et al.~\cite{Radicchi:2004};
(ii) spectral graph partitioning~\cite{Capocci:2004}; (iii)
agglomerative methods, including hierarchical
clustering~\cite{Scott1979,wasserman94}; (iv) maximization of the
modularity, as in Newman~\cite{Newman:2003} and Duch and
Arenas~\cite{Duch:2005}; and (v) consideration of successive
neighborhoods through hierarchical growth emanating from
hubs~\cite{Costa:2004,Bagrow05}. A good survey of community
identification methods has been provided by Newman
\cite{Newman:2004:EPJ} and Danon et al. \cite{Danon:2005}.
This subject has also been partially addressed in the surveys by Costa
et al.~\cite{Costa:survey} and Boccaletti et al~\cite{Boccaletti05}.

Arguably, the most popular method for community identification is that
proposed by Girvan and Newman~\cite{Girvan:2003}.  This approach
considers that the edges interconnecting communities correspond to
bottlenecks between the communities, so that the removal of such edges
tend to partition the network into communities. The bottleneck edges
are identified in terms of a measurement called \emph{edge
betweenness}, which is given by the number of shortest paths between
pairs of vertices that run along the edge.  This algorithm has been
proven to be effective for obtaining communities in several types of
networks. However, its effectiveness implies a computational cost of
order $O(n^2m)$ in a network with $m$ edges and $n$ vertices.  An
alternative algorithm to calculate betweenness centrality, based on
random walks, has been proposed~\cite{Newman:2004:PRE} which, although
conceptually interesting, is also computationally demanding.

The method described in the present article overcomes tends to run
faster than the Girvan-Newman's algorithm while offering reasonable,
though smaller, precision for identification of communities.  It is
based on the consideration of successive neighborhoods of a set of
seeds, implemented through hierarchical growth.  Starting from a
vertex (seed), the links of its successive neighborhood are analyzed
in order to verify if they belong to the same community than the
seed. This process starts from each vertex in the network and, at each
step, inter-community edges are removed splitting the network into
communities.

A related approach was previously proposed by Costa~\cite{Costa:2004},
who developed a method based on the flooding the network with
wavefronts of labels emanating simultaneously from hubs. The expanding
region of each label was implemented in terms of hierarchical growth
from the starting hubs and the communities are found when the
wavefronts of labels touch each one.  Competitions along the
propagating neighborhoods are decided by considering an additional
criterion involving the moda of the labels at the border of the
neighborhood and the number of emanating connections.  The possibility
to detect communities by using expanding neighborhoods has also been
addressed by Bagrow and Bollt~\cite{Bagrow05}, who proposed an
algorithm based on the growth of an \emph{l-shell} starting from a
vertex $v_0$, with the process stopping whenever the rate of expansion
is found to fall bellow an arbitrary threshold. The
\emph{l-shell} is composed by a set of vertices placed at distance
$l$ from the vertex $v_0$, which is analogous to the concept of
\emph{ring} defined by Costa~\cite{Costa:2005,Costa05b} in order
to introduce hierarchical measurements. At each expansion, the total
emerging degree of a shell of depth $l$ is calculated as corresponding
to the sum of the emerging degree of each vertex at distance $l$ from
$v_0$, i. e. the degree of $i$ minus the number of links that connect
$i$ with vertices inside the shell (analogous to the concept of
hierarchical degree introduced by
Costa~\cite{Costa:2005,Costa05b}). When the rate between the total
emerging degree at distance $l$ and $l-1$ is shorter than a given
threshold, the set of vertices inside the \emph{l-shell} is classified
as a community. Despite its simplicity, the determination of the local
community is accurate just when the vertex $v_0$ is equidistant from
all parts of its enclosing community~\cite{Bagrow05}. In order to
overcome this limitation, Bragrow and Bollt suggested starting from
each vertex and then find a consensus partitioning of the network
using a membership matrix.  Such an approach makes the algorithm more
precise.  On the other hand, it is slow because it requires sorting
the membership matrix, which is of order $O(n^3)$.

The method reported in the present article also involves the
consideration of expanding neighborhoods and completion of growth in
terms of rate of expansion. However, it differs from the method of
Bagrow and Bollt because it analyzes the connections of each vertex at
the border of the community individually instead of all vertices at
same time. Besides, it considers not only the first neighborhood of
the community, but the second one too. At each expansion from an
starting vertices, edges can be removed considering two trials based
on the first and second neighborhood of the enclosing
community. Another difference is that our method uses a threshold just
at the second neighborhood, whose value is determined so as to obtain
the best value of the modularity, i. e.  the value of this threshold
varies from $0$ to a maximum value and at each variation it is
computed the modularity. The procedure is to that used by
Girvan-Newman, as the modularity is calculated at each edge removal.

The next sections describe the suggested method as well as its
application to community detection in real and in computer generated
networks.  A comparison with the Girvan-Newman method in terms of
precision and execution time is also presented and discussed.

\section{Hierarchical growth method}

A community is formed by a set of densely connected vertices which is
sparsely connected with the remainder of the network. The proposed
hierarchical growth method finds communities by considering two
expanding neighborhoods. The first neighborhood of a given vertex is
composed by those vertices at a distance of one edge from that
vertex. Similarly, the set of vertices at distance of two edges from
that given vertex constitutes its second neighborhood.  Following this
definition, two steps are performed in order to determine if a given
vertex $i$ located in the first neighborhood of a known community
belongs to this community, i.e.

\begin{enumerate}
    \item \begin{equation}\label{cond1} \frac{k_{in_1}(i)}{k_{out_1}(i)} \geq 1,\end{equation}
    where $k_{in_1}(i)$ is the number of links of the vertex $i$ with
    vertices belonging to community and with vertices in the first
    neighborhood, and $k_{out_1}(i)$ is the number of links between
    the vertex $i$ and vertices in the remainder of the network.
    \item
    \begin{equation}\label{cond2}
         \frac{k_{in_2}(i)}{k_{out_2}(i)} >\alpha,
    \end{equation} 
    where $k_{in_2}(i)$ is the number of links of
    the neighbors of $i$ located in the second community neighborhood
    with vertices belonging to the first neighborhood, and
    $k_{out_2}(i)$ is the number of links between the neighbors of $i$
    and vertices in the remainder of the network.  The parameter
    $\alpha$ varies from $1$ to a threshold value which is determined
    according to the higher value of the modularity.
\end{enumerate}

The first condition is sufficient to determine if a vertex belongs to
the community, but it is not necessary. The coefficient $\alpha$ acts
as a threshold ranging from one to a maximum value. The extension of
the current method for weighted network is straightforward.

\begin{figure}
\begin{center}
  \centerline{\includegraphics[width=8cm]{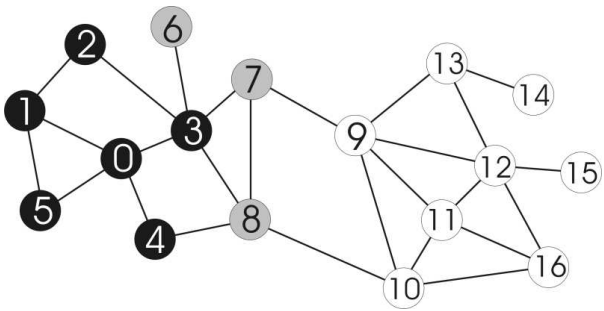}}
  \caption{Application example of the hiearchical growth. The process
  is started at vertex $0$. Its neighborhood, indicated by black
  vertices, are analyzed next, and the vertices $1,2,4$ and $5$ are
  added to community following the first condition (equation
  \ref{cond1}).  The vertex $3$ is added to community following the
  second condition (equation \ref{cond2} with $\alpha = 1$). The
  current community neighborhood (gray vertices) is then checked, and
  the vertices $6, 7$ and $8$ are added because of the first
  condition. Next, the links between the community and the vertices
  $9$ and $10$ are removed, splitting the network into two
  communities.  } \label{example}
\end{center}
\end{figure}

The hierarchical growth starts from each vertex of the network at each
step, with the vertices with highest clustering
coefficient~\cite{Costa:survey} selected first because they are more
likely to be inside communities. So, the first and/or the second
conditions are analyzed at each step, while the ring between the
starting vertex grows, adding vertices to the community or removing
edges. Nodes satisfying the first and/or the second
conditions~(equations~\ref{cond1} and~\ref{cond2}) are added to the
community. Otherwise, their links with the known community are
removed. Figure~\ref{example} illustrates a simple application example
of the method. In order to determine the best division of the network
the threshold $\alpha$ is varied from $0$ to a maximum value and at
each variation, the modularity $Q$ is computed. The modularity is a
measure of the quality of a particular division of
networks~\cite{Newman:2004:PRE}. If a particular network is to be
split in $c$ communities, $Q$ is computed defining a symmetric
$c\times c$ matrix $E$ whose elements of diagonal, $e_{ii}$, give the
connections between vertices in the same community and the remainder
elements, $e_{ij}$, give the number of connections between the
communities $i$ and $j$,

\begin{equation}\label{modularity}
Q = \sum_i [ e_{ii} - (\sum_j e_{ij})^2 ] = Tr(E) - ||E^2||,
\end{equation}

where $Tr(E)$ is the trace of matrix $E$ and $||E||$ indicates the
sum of the elements of the matrix $E$.

Thus, the splitting of the network considers the value of $\alpha$
that provides the highest value of the modularity. The pseudocode
which describes the hierarchical growth method is given in Algorithm
1.

\begin{algorithm}[t]
\label{alg} \caption{The general algorithm for the hierarchical
growth method.}
\begin{algorithmic}%[1]
\FOR{each vertex of the network}
    \STATE put the next vertex with highest clustering coefficient value in $\mathcal{C}$
    \WHILE{$\mathcal{C}$ does not stop growing}
        \STATE put the neighbors of $\mathcal{C}$ in $\mathcal{R}$
        \FOR{each vertex $i$ in $\mathcal{R}$}
            \STATE compute $k_{in_1}(i)$ and $k_{out_1}(i)$
            \IF{$k_{in_1}(i) \geq k_{out_1}(i)$}
                \STATE insert the vertex $i$ in $\mathcal{C}$
            \ELSE
                \STATE select the neighbors of $\mathcal{R}$ and put in $\mathcal{R}1$
                \STATE compute $k_{in_2}(i)$ and $k_{out_2}(i)$
                \IF{$k_{in_2}(i) > \alpha k_{out_2}(i)$}
                    \STATE insert the vertex $i$ in $\mathcal{C}$
                \ELSE
                    \STATE remove the links between the vertex $i$ and the vertices in $\mathcal{C}$
                \ENDIF
            \ENDIF
        \ENDFOR
    \ENDWHILE
    \STATE Clean $\mathcal{C}, \mathcal{R}$ and $\mathcal{R}1$
\ENDFOR
\end{algorithmic}
\end{algorithm}

\section{Applications}

In this section we illustrate applications of the hierarchical growth
to particular problems while analyzing its accuracy and the
performance. In the first case, its accuracy is determined by
comparing the obtained results with expected divisions of different
networks. With the purpose of determining the performance, we compared
the hierarchical growth method with Girvan-Newman's algorithm, whose
implementation is based on the algorithm developed by
Brandes~\cite{Brandes01} for computing of vertex betweenness
centrality.

In order to split the network into communities the Girvan-Newman
algorithm proceeds as follows:
\begin{enumerate}
    \item Calculate the betweenness score for each of the edges.
    \item Remove the edge with the highest score.
    \item Compute the modularity for the network.
    \item Go back to step 1 until all edges of the
    networks are removed, resulting in $N$ non-connected nodes.
\end{enumerate}

The best division is achieved when the highest modularity value is
obtained. In this way, the Girvan-Newman method runs in two steps: (i)
first all edges are removed from the network and the modularity value
is computed at each removal, (ii) next, the highest value of
modularity is determined and the corresponding edges removed.

\subsection{Computer generated networks}

A typical procedure to quantify how well a community identification
method performs adopts networks with known community structure, called
\emph{computer generated networks}, which are constructed by using two
different probabilities~\cite{Newman:2004:PRE}. Initially, a set of
$n$ vertices are classified into $c$ communities. At each subsequent
step, two vertices are selected and linked with probability $p_{in}$
if they are in the same community, or $p_{out}$ in case they are
belong to different communities. The values of $p_{in}$ and $p_{out}$
can be selected so as to control the sharpness of the separation
between the communities. When $p_{in}~\ll~p_{out}$, the communities
can easily be visualized. On the other hand, when $p_{in} \rightarrow
p_{out}$, it is difficult to distinguish the communities and the
methods used for community identification lose precision in the
correct classification of the vertices into communities.

\begin{figure}[ht]
\begin{center}
  \centerline{\includegraphics[width=8.5cm]{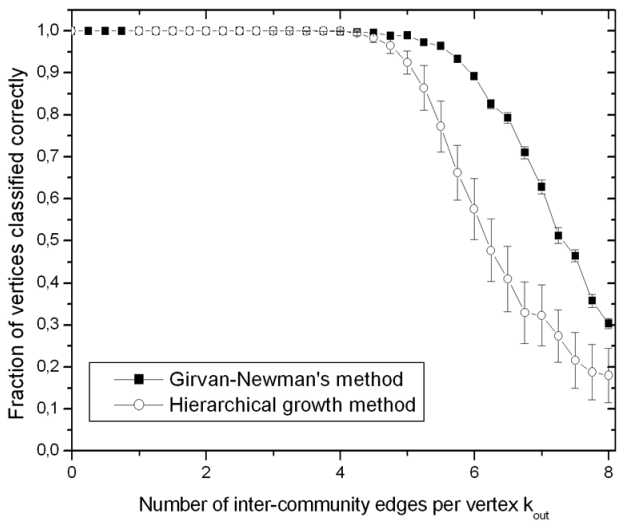}}
  \caption{Fraction of correctly classified vertices in terms of the
  number of inter-community edges $k_{out}$ for a network with $128$
  vertices considering $k_{in}+k_{out}=16$.  The
  Girvan-Newman's method is more precise than the hierarchical growth method when
  $k_{out}>5$.  Each data point is an average over $100$ graphs.}
 \label{precision}
\end{center}
\end{figure}

We generated networks with $128$ vertices, divided into four
communities of $32$ vertices each. The total average vertex degree
$k_{in} + k_{out}$ of the network was kept constant and equal to
$16$. In this way, as the value of $k_{out}$ is varied from $0$ to
$8$, the more difficult the network communities recognition
becomes. The proposed community finding algorithm was applied to each
network configuration, and the fraction of vertices classified
correctly was calculated. In Figure~\ref{precision} it is shown the
sensitivity of the hierarchical growth method compared with the
results obtained by using Girvan-Newman's method.

As Figure~\ref{precision} shows, the algorithm performs near full
accuracy when $k_{out}\leq5$, classifying more than $90\%$ of
vertices correctly. For higher values, this fraction falls off as
the connections between communities gets denser. When
$k_{out}>5$, the Girvan-Newman's method gives a better result,
so it tends to be more suitable for this kind of networks.

\begin{figure}[ht]
\begin{center}
  \centerline{\includegraphics[width=8.5cm]{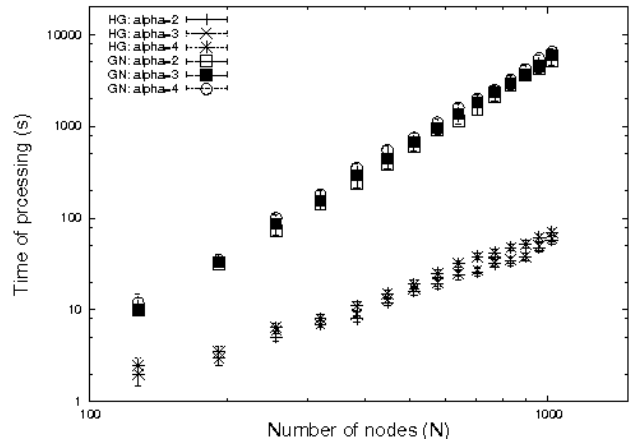}}
  \caption{Processing time versus the size of network. The
  hierarchical growth (HG) method runs faster than the Girvan-Newman
  (GN) method. While the time of processing of the Girvan-Newman's
  method scales as $N^{3.0\pm 0.1}$, the time of hierarchical growth
  method scales as~$N^{1.6\pm 0.1}$. Each data point is an average
  over 10 graphs.}  \label{time}
\end{center}
\end{figure}

The execution times of both methods were compared considering the
computer generated cases for which the hierarchical growth method
provides exact results (i.e. we used $kout=2,3$ and $4$). We
considered the network size varying from $N = 128$ until $N = 1,024$
and kept the average degree $k_{in}+k_{out}=16$. The hierarchical
growth method resulted faster than the Girvan-Newman's method, as
shown in Figure~\ref{time}. While the Girvan-Newman's processing time
scales as $N^{3.0\pm0.1}$, the time of the hierarchical growth method
scales as $N^{1.6\pm0.1}$, which suggests that the former method is
particularly suitable for large networks.

The constant $\alpha$ considered in the algorithm is determined in the
following way. The algorithm runs for $\alpha$ varying from $1$ to a
maximum value $\alpha_M$ increasing in steps of $0.5$. For each value
of $\alpha$, the communities are computed, and the decomposition with
the best value of modularity is chosen. In our tests, the best value
of $\alpha$ was always equal to $1$ for all network sizes considered.

\subsection{Zachary karate club network}

\begin{figure}[ht]
\begin{center}
  \centerline{\includegraphics[width=8.5cm]{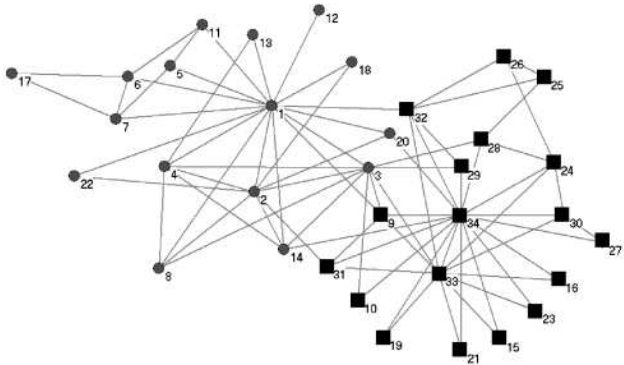}}
  \caption{The friendship Zachary karate club network divided into two
  communities, represented by circles and squares. The division
  obtained by the hierarchical growth is the same as the one provided
  by the Girvan-Newman's method.}  \label{zachary}
\end{center}
\end{figure}

In order to apply the hierarchical growth method to a real network, we
used the popular Zachary karate club network~\cite{Zachary:1977},
which is considered as a simple benchmark for community finding
methodologies~\cite{Newman:2004:EPJ,Duch:2005,Costa:2004}. This
network was constructed with the data collected observing $34$ members
of a karate club over a period of $2$ years and considering friendship
between members. The two obtained communities are shown in
Figure~\ref{zachary}. This partitioning of the network corresponds
almost perfectly to the actual division of the club members, while
only one vertex, i.e. vertex $3$, has been misclassified. This result
is analogous to that obtained by using the Girvan-Newman algorithm
based on measuring of betweenness centrality~\cite{Girvan:2003}.

\subsection{Image segmentation}

A third application of our method is related to the important problem
of image segmentation, i.e. the partition of image elements (i.e
pixels) into meaningful areas corresponding to existing objects.  As
described by Costa~\cite{Costa04-vision}, an image can be modeled as a
network and methods applied to networks characterization can be used
to identify image properties. The application of a community finding
algorithm to image segmentation was proposed in that same
work~\cite{Costa:2004}.  Since digital images are normally represented
in terms of matrices, where each element corresponds to a pixel, it is
possible to associate each pixel to a node using network image
representation. The edge weight between every pair of pixels can be
determined by the Euclidean distance between feature vectors composed
by visual properties (e.g. gray-level, color or texture) at or around
each pixel. Thus, considering the distance between every feature
vector of pair of pixels in the image, this approach results in a
fully-connected network, where closer pixels are linked by edges with
higher weights. To eliminate weak links, a threshold can be adopted
over the weighted network, resulting in a simplified adjacency
matrix. The connections whose distance is shorter than the threshold
are assigned to zero, otherwise, to one.

\begin{figure}[t]
\begin{center}
  \centerline{\includegraphics[width=8.50cm]{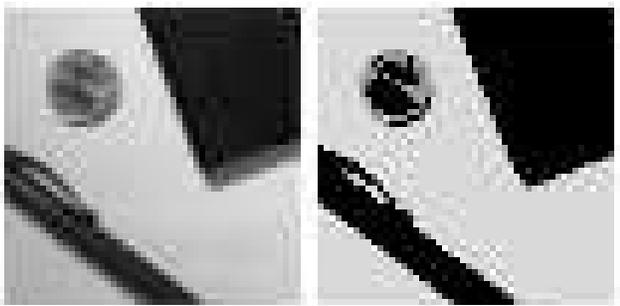}}
  \caption{The real image and its respective segmentation. The image
  is transformed into a network and a threshold $T = 0.25$ is used to
  eliminate weak links.}
 \label{segmentation}
\end{center}
\end{figure}

The mapping between a pixel in the image to a node in the network
and the reverse operation, is defined~\cite{Costa:2004} by
\begin{eqnarray}
  i = y + (x-1)M, \label{eq:map1}  \\
  x = \lfloor (i-1)/M \rfloor + 1, \label{eq:map2}  \\
  y = mod((i-1),M) + 1, \label{eq:map3}
\end{eqnarray}

where $M$ is the size of the square image, and $1\leq~x,y~\leq~M$ are
the pixel positions in the image. In this way, the resulting weighted
network has $N = M^2$ nodes and $n = N(N-1)/2$ edges.

Figure~\ref{segmentation} shows the initial image and its respective
segmentation. The results obtained by the hierarchical growth method
and by using the Girvan-Newman's method are similar. Since the network
obtained typically for images can be substantially large ($N = M^2$),
a faster method to community identification is necessary for practial
applications, a demand potentially met by hierarchical growth method.

\section{Conclusions}

In this paper we have proposed a new method to identify communities in
networks. The method is based on a hierarchical growth from a starting
node while its neighborhood is analyzed, and edges removed according
to two rules based on the first and/or second neighborhoods of the
growing community. We have applied this method to computer generated
networks in order to determine its precision and performance comparing
it with the popular method based on edge betweenness centrality
proposed by Girvan and Newman~\cite{Girvan:2003}. Despite resulting
not so precise as the Girvan-Newman's method, the proposed algorithm
is promisingly fast for determining communities.  We have also applied
the hierarchical growth method to the Zachary karate club network and
image segmentation. In both cases, the resulting networks are similar
to those obtained by the Girvan-Newman's algorithm.

As discussed by Danon et al.~\cite{Danon:2005}, the most accurate
methods tend to be computationally more expensive. The method
presented in this article can not provide as good precision as most of
the methods, but it yields competing velocity. As a matter of fact,
performance and accuracy need to be considered when choosing a method
for practical purposes. Particularly in the case of image
segmentation, the suggested method is particularly suitable given the
large size of the typical networks (increasing with the square of the
image size, $N = M^2$) and the sharped modular structure often found
in images.

As a future work, the algorithm proposed here can be improved
considering other conditions to include nodes in the growing community
as, for example, higher levels of community neighborhood. Besides,
consideration of local modularity can be also considered in order to
obtain a more precise partition of the network.

\section{Acknowledgments}

Luciano da F. Costa is grateful to FAPESP (proc. 99/12765-2), CNPq
(proc. 308231/03-1) and the Human Frontier Science Program
(RGP39/2002) for financial support. Francisco A. Rodrigues
acknowledges FAPESP sponsorship (proc. 04/00492-1).

\bibliographystyle{unsrt}
\bibliography{hierarchical_growth}

\end{document}